\documentclass[twocolumn]{aastex631}
\usepackage{subfigure}
\usepackage{multirow} 
\usepackage{amsmath}

\received{June 25, 2024}
\revised{August 27, 2024}
\accepted{September 15, 2024}

\begin{document}

\title{Estimating the total energy content in escaping accelerated solar electron beams}

\correspondingauthor{Alexander W. James}
\email{alexander.james@ucl.ac.uk}

\author[0000-0001-7927-9291]{Alexander W. James}
\affiliation{Mullard Space Science Laboratory, University College London \\
Holmbury St. Mary, Dorking \\
 Surrey, RH5 6NT, UK}

\author[0000-0002-6287-3494]{Hamish A. S. Reid}
\affiliation{Mullard Space Science Laboratory, University College London \\
Holmbury St. Mary, Dorking \\
 Surrey, RH5 6NT, UK}

\begin{abstract} 
Quantifying the energy content of accelerated electron beams during solar eruptive events is a key outstanding objective that must be constrained to refine particle acceleration models and understand the electron component of space weather.  Previous estimations have used in situ measurements near the Earth, and consequently suffer from electron beam propagation effects.
In this study, we deduce properties of a rapid sequence of escaping electron beams that were accelerated during a solar flare on 22 May 2013 and produced type III radio bursts, including the first estimate of energy density from remote sensing observations.
We use extreme-ultraviolet observations to infer the magnetic structure of the source active region NOAA 11745, and Nan\c{c}ay Radioheliograph imaging spectroscopy to estimate the speed and origin of the escaping electron beams. Using the observationally deduced electron beam properties from the type III bursts and co-temporal hard X-rays, we simulate electron beam properties to estimate the electron number density and energy in the acceleration region.
We find an electron density (above $30\ \mathrm{keV}$) in the acceleration region of $10^{2.5}\ \mathrm{cm}^{-3}$ and an energy density of $2\times10^{-5}\ \mathrm{erg\ cm}^{-3}$. Radio observations suggest the particles travelled a very short distance before they began to produce radio emission, implying a radially narrow acceleration region. A short but plausibly wide slab-like acceleration volume of $10^{26}-10^{28}\ \mathrm{cm}^{3}$ atop the flaring loop arcade could contain a total energy of $10^{23}-10^{25}\ \mathrm{erg}$ ($\sim 100$ beams), which is comparable to energy estimates from previous studies.
\end{abstract}


\section{Introduction} \label{sec:intro}

The Sun is the most prolific particle accelerator within the solar system and our close proximity provides one of the best opportunities to analyse and understand astrophysical particle acceleration and transport. Exactly how much energy goes into accelerating electrons during solar eruptive events is part of a major outstanding science question in this field. Such information is crucial for refining particle acceleration models and to understand the electron component of space weather.

\citet{krucker2007electron} calculated the number of electrons needed to produce observed hard X-rays, assuming the thick target model \citep{brown1971thick}, and compared this to the number of electrons that escaped into interplanetary space, as inferred from electron spectra measured at 1 AU. Whilst they found the hard X-ray sources were typically produced by $10^{35}\textrm{--}10^{36}$ electrons above $50\, \mathrm{keV}$, only $10^{31}\textrm{--}10^{33}$ electrons above the same energy escaped (an average of $\sim 0.2\%$ escaping electrons).

Where \citet{krucker2007electron} examined 16 relatively strong flares (9 M-flares, 6 C-flares and 1 B-flare), \citet{james2017electron} performed a similar analysis using six weak ($\leq \mathrm{C}1.0$) flares.
Also assuming the thick target model (which is supported by their observations), \citet{james2017electron} found their hard X-ray sources were produced by $10^{30}\textrm{--}10^{33}$ electrons, whilst $10^{30}\textrm{--}10^{32}$ electrons escaped. 
Interestingly, the fraction of electrons that produced hard X-ray sources and electrons that escaped varied significantly from flare to flare. 
In one flare, the number of escaping electrons was $\approx 6\%$ of the number that produced the hard X-ray sources, but in another, the same fraction was $148\%$, \textit{i.e.} more electrons escaped than produced the hard X-ray emission. 
Furthermore, by fitting the differential electron flux measured at 1 AU, \citet{james2017electron} found energies in the escaped electron beams of $\approx \num{e23}\textrm{--}\num{e25}\ \mathrm{erg}$ above the break energy of $74\, \mathrm{keV}$. The energies of the associated hard X-ray electron signatures at the Sun ranged from $10^{24}\textrm{--}10^{26}\, \mathrm{erg}$.

For another comparison, \citet{dresing2021electron} studied 17 B-class and C-class flares (somewhat intermediate between the weak flares studied by \citealt{james2017electron} and the strong flares of \citealt{krucker2007electron}). \citet{dresing2021electron} identified a break energy of $45\, \mathrm{keV}$ (close to that seen by \citealt{krucker2007electron}) and found $10^{33}\textrm{--}10^{34}$ electrons in the hard X-ray flares and $10^{30}\textrm{--}10^{31}$ electrons in the escaping beams above the break energy. This corresponds to low fractions of escaping electrons, ranging from $0.18\textrm{--}0.24\%$, supporting the work on stronger flares by \citet{krucker2007electron}.

To estimate the energy content of escaping energetic electrons using remote sensing observations, instead of in situ measurements, we require estimations of three quantities of the accelerated electrons: the energy distribution, the volume of the source and the number density.

The energy distribution of escaping electrons is measured as a power-law, typically with a spectral break that can be situated either in the deca-keV range or up to a few 100s keV \citep{Krucker2009}.  The spectral break has been associated with the generation of Langmuir waves during propagation through simulations \citep{KontarReid2009,reid2010electrons, reid2013electrons} and via observations \citep{Lorfing2023Langmuir}, and also the presence of pitch-angle scattering \citep{dresing2021electron}.  The three studies detailed above \citep{krucker2007electron,james2017electron,dresing2021electron} found a very good agreement with the spectral index of the power-law electron energy distribution derived from X-rays, and the spectral index of the electron beam power-law energy distribution measured in situ, above the break energy.

Many previous studies have supported the idea that these hard X-ray emissions and solar radio bursts are products from common particle acceleration events. The emissions are often correlated in time (both their peak emissions and short-timescale variations) and space, with spatially-resolved imaging demonstrating radio sources close to flaring regions with hard X-ray sources (for a comprehensive overview, see \citealt{PickVilmer2008,reid2014starting,reid2017radioXray} and the many references within).  Some radio bursts and hard X-ray sources that first appear to be linked may originate from subsequent episodes of reconnection in different locations, suggesting multiple episodes/stages of reconnection and particle acceleration \citep[\textit{e.g.}][]{vilmer2003resolved}.  There was also less agreement in the spectral index when in situ electrons were delayed compared to the onset of hard X-rays \citep{krucker2007electron}.  However, the general agreement between hard X-rays, radio bursts and in situ energetic electrons provides reasonable confidence for estimating the energy distribution of escaping electron beams using X-ray measurements.  

The volume of solar acceleration regions are currently ill-defined, largely due to the lack of direct electromagnetic emission.  It is not until electrons propagate away from the acceleration region that they typically produce hard X-rays or solar radio bursts.   In the standard 2D CSHKP model of solar flares \citep{carmichael1964flare,sturrock1966flare,hirayama1974flare,kopp_pneuman1976reconnection}, the acceleration of particles occurs at an `X-point' beneath an overlying rising plasma structure (\textit{e.g.} an erupting magnetic flux rope) and above the observable flaring arcade. Extending this concept to 3D, this `X-point' becomes a quasi-separatrix layer (QSL; \citealp{priest1995qsl,demoulin1996qsl}), manifesting as a current sheet which separates the flare arcade and the overlying structure \citep{janvier2013standard}. 

Previously, \citet{guo2012accelregion} suggested the loop top acceleration region studied in their work was of the order half the width of the observed loops, corresponding to 10--15 Mm. 
Furthermore, \citet{gordovskyy2020acceleration} identified a cylindrical acceleration region with length of $20\textrm{--}25\, \mathrm{Mm}$ (of similar order to \citealt{guo2012accelregion}) and a $5\, \mathrm{Mm}$ diameter, equating to a volume of $10^{26}\, \mathrm{cm}^{3}$.
Whilst not necessarily exclusively representative of the acceleration region, \citet{fleishman2022volume} estimated that a loop top region of interest where they saw the main energy release in a flare had a spherical volume of $10^{27}\, \mathrm{cm}^{3}$. 

Beams of electrons cause type III radio bursts, which are characterised by starting frequencies of $<1\ \mathrm{GHz}$ \citep[][and references within]{reid2014review}, suggesting they originate from regions of the corona with higher densities (typically, lower altitudes), and fast frequency drift rates as the high-speed electrons rapidly traverse the coronal plasma density gradient.  A remarkably small acceleration region was identified by \citet{chen2018beams} using GHz radio emission detected using the VLA, with a plane of sky area of only $600\, \mathrm{km}^{2}$.  Using lower frequency type III radio bursts, the longitudinal extent of acceleration regions was deduced from combined X-ray and radio observations, relating to the low-high-low trend in starting frequency that mirrors the soft-hard-soft trend in X-ray spectral index.  Considering a beam of electrons that travels radially away from the Sun, the coronal height at which type III radio bursts begin, $h_{\mathrm{III}}$ is not the location where the electron beams were accelerated, $h_{\mathrm{acc}}$. Rather, these heights are separated by the instability distance, $d\alpha$
\begin{equation}
    h_{\mathrm{III}} = h_{\mathrm{acc}} + d\alpha,
    \label{eqn:instability_distance}
\end{equation}
where $\alpha$ is the electron spectral index and $d$ is the length of the acceleration region \citep{reid2011acceleration,reid2014starting}.  Therefore, the length of an acceleration region, $d$, can be quantified by measuring the associated spectral index, $\alpha$, observing the height at which type III radio bursts begin, and deducing the location of the acceleration region.  \citet{reid2014starting} used this technique to deduce acceleration region lengths of 2--13 Mm.

Approximations for the cross-sectional area of electron beams can be deduced from the area of radio burst sources.  The full-width half-maximum (FWHM) radio source size was shown by \citet{kontar2019scattering} to vary with frequency, $f$, close to $f^{-1}$ using a combination of observations from numerous studies.  One can extrapolate radio source size down to 1 GHz, providing a FWHM around 30 Mm.  However, we might expect the actual source sizes to be slightly smaller as scattering from density fluctuations can artificially increase our measured source sizes, particularly for fundamental radio emission, where the frequency of radio waves are emitted close to the plasma frequency.

The number density of escaping electrons is not typically estimated from remote sensing observations.  The nonlinear plasma emission mechanism makes deducing electron beam densities from radio bursts difficult.  A connection between the electron energy density and the bulk electron beam velocity deduced from radio bursts has been found numerically \citep{reid_kontar2018}.  Therefore, by measuring the spectral index and estimating the beam velocity, we can estimate the electron number density in the solar acceleration region.  We can compare this electron density to that observed in hard X-ray sources that are produced when the downward component of the beam bombards the chromosphere to gain insight in to the fraction of particles that are accelerated.  From this, we can quantify the energy in the electron beam, which is useful for understanding space weather effects.

In this work, we analyse observations of type III radio burst activity that occurred around the time of a CME and an associated M-class flare on 22 May 2013, simulate one of the electron beams to determine the electron density at the beam source, and use the derived energetics of the event to infer the size and location of the particle acceleration region.
We outline the data used in this work in Section \ref{sec:data} and present extreme-ultraviolet (EUV) observations of a CME and flare in Section \ref{sec:observations}. We analyse X-ray and radio observations of accelerated particle beams in Sections \ref{sec:confined_beams} and \ref{sec:radio}, and describe our simulation of an observed electron beam in Section \ref{sec:simulation}. Finally, we discuss our findings in Section \ref{sec:discussion} and summarise the main conclusions in Section \ref{sec:conclusions}.

\section{Data} \label{sec:data}

We use radio images taken by the \textit{Nan\c{c}ay Radioheliograph} (NRH; \citealp{kerdraon1997NRH}) at nine frequencies between $150.7\ \mathrm{MHz}$ and $445.0\ \mathrm{MHz}$. The images span $4\, \mathrm{R}_{\odot}$ with a spatial resolution of 30$\arcsec{}$ and a cadence of 0.225 seconds. 
For added context, we also look at radio emission in the range $20\ \mathrm{kHz}-1004\ \mathrm{MHz}$ using the Wind/WAVES ($20\ \mathrm{kHz}\textrm{--}13.825\ \mathrm{MHz}$; 
\citealp{bougeret1995WAVES,ogilvie1997Wind}), NDA ($10\textrm{--}80\ \mathrm{MHz}$; \citealp{Lecacheux2000NDA}) and ORFEES ($144\textrm{--}1004\ \mathrm{MHz}$; \citealp{hamini2021ORFEES}) instruments.

We use hard X-ray images and spectra from the \textit{Reuven Ramaty High-Energy Solar Spectroscopic Imager} (RHESSI; \citealp{lin2002rhessi}). 
Hard X-ray fluxes were analysed in the range $3\ \mathrm{keV} \textrm{--} 200\ \mathrm{keV}$.
We also examine the full-disc integrated soft X-ray intensity of the Sun measured by the \textit{Geosynchronous Operational Environmental Satellite} (GOES) network in the $1.0\textrm{--}8.0$ \AA{} range.

EUV observations of the corona are made by the \textit{Atmospheric Imaging Assembly} (AIA; \citealp{lemen2012atmospheric}) onboard the \textit{Solar Dynamics Observatory} (SDO; \citealp{pesnell2012SDO}).
We examined data from all channels of AIA, but particularly of note in this paper are observations from the 171 \AA{} channel (temperature response peak at $10^{5.8}\ \mathrm{K}$), the 193 \AA{} channel (double-peaked at $10^{6.2}\ \mathrm{K}$ and $10^{7.3}\ \mathrm{K}$), 211 \AA{} ($10^{6.3}\ \mathrm{K}$), 131 \AA{} (response peaks at $10^{5.6}\ \mathrm{K}$ and $10^{7.0}\ \mathrm{K}$) and 1600 \AA{} ($10^{5.0}\ \mathrm{K}$). 
We also use additional EUV observations taken by the the \textit{Extreme Ultraviolet Imager} (EUVI) onboard the \textit{Solar Terrestrial Relations Observatory A} (STEREO A; \citealp{kaiser2008stereo}) spacecraft for a complementary view of the western limb $137^{\circ}$ ahead of the Earth.

We identify CMEs using white light observations from the \textit{Large Angle and Spectrometric Coronagraph} (LASCO; \citealp{brueckner1995large}) onboard the \textit{Solar and Heliospheric Observatory} (SOHO; \citealp{domingo1995soho}). Regions of coronal magnetic field that were open to the heliosphere are found using potential field source surface (PFSS) extrapolations of photospheric magnetograms made in IDL SolarSoft.

\section{CME and Flare Observations} \label{sec:observations}

\subsection{The CME}
 
Two CMEs erupted from the Sun's western limb (as seen from Earth) on 22 May 2013. The first CME entered the LASCO C2 field of view at 08:48 UT and the second at 13:25 UT (times from the SOHO LASCO CME catalog\footnote{\url{https://cdaw.gsfc.nasa.gov/CME_list/}}). Previous studies have estimated the speeds of these CMEs using the Graduated Cylindrical Shell method \citep{thernisien2006GCS,thernisien2009GCS,thernisien2011GCS}, finding speeds of $\approx\! 500\ \mathrm{km\ s}^{-1}$ and $\approx\! 1500\ \mathrm{km\ s}^{-1}$, respectively \citep{ding2014interaction,palmerio2019ApJmultipoint}. 

 \begin{figure*}[!htb]
 \centerline{\includegraphics[width=1.0\textwidth,clip=]{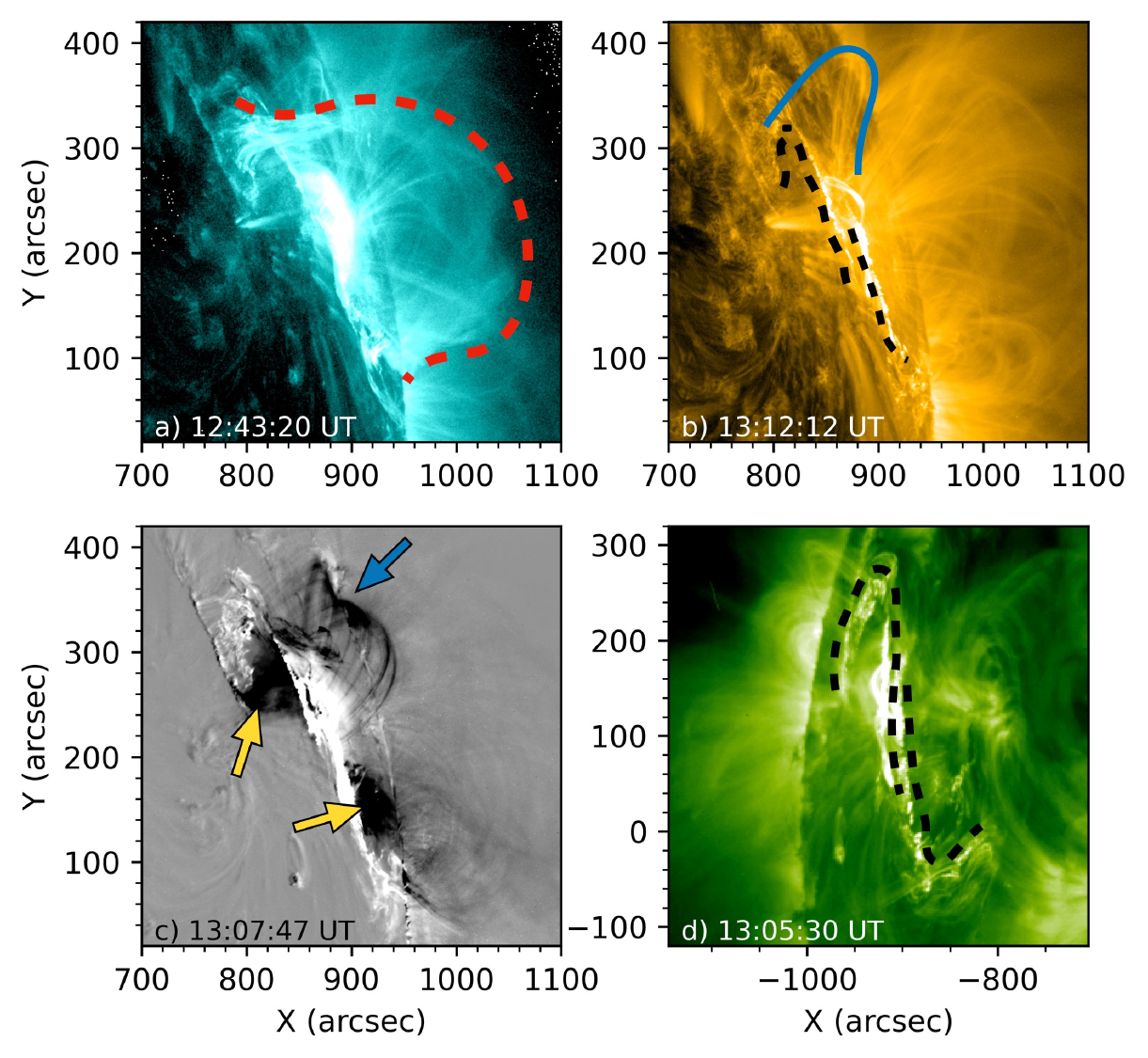}}
 \caption{a) Arched plasma structure seen in the AIA 131 \AA{} channel (outlined with a red dashed line), interpreted as a hot magnetic flux rope signature b) Overlying loops (blue line) and two J-shaped flare ribbons (black dashed lines) seen in the AIA 171 \AA{} channel. c) Twin EUV dimmings seen in base difference images in the AIA 211 \AA{} channel (yellow arrows, base time at 12:00 UT on 22 May 2013). Overlying loops (blue arrow) dim and disappear from in this channel as they are deflected by the erupting CME. d) Two J-shaped flare ribbons seen in the 195 \AA{} channel of STEREO-A/EUVI, outlined with black dashed lines.}
 \label{fig:aia_multi}
 \end{figure*}

Extreme-ultraviolet observations from SDO/AIA show that the second CME originated from NOAA AR 11745, which was at the western limb in the northern solar hemisphere. At 12:17 UT, filament material in the active region brightened and an arched plasma structure was seen to rise until it left the AIA field of view at $\approx$13:00 UT. This arched structure was observed in the 131 \AA{} channel of AIA (see Figure \ref{fig:aia_multi}a) but not in the 171 \AA{} channel (see Figure \ref{fig:aia_multi}b) demonstrating that the emitting plasma was very hot at around $10^{7.0}\ \mathrm{K}$. We suggest this structure corresponds to an erupting magnetic flux rope, as with other similar hot plasma emission structures seen in the AIA 131 \AA{} channel \citep[\textit{e.g.},][]{cheng2011rope,zhang2012rope,patsourakos2013rope,nindos2015ropes,james2017on-disc,james2020trigger}. 

Further evidence of an erupting flux rope comes from observed double-J hooked flare ribbons \citep[\textit{e.g.}][]{janvier2013standard} that brighten during the eruption, particularly in the AIA 171 \AA{} and STEREO A 195 \AA{} channels (see Figure \ref{fig:aia_multi}b and \ref{fig:aia_multi}d), and twin EUV dimmings seen clearly in base difference images of the AIA 211 \AA{} channel (Figure \ref{fig:aia_multi}c).

Overlying loops (AIA 193 \AA{} and 171 \AA{}; Figure \ref{fig:aia_multi}b) are seen to deflect away from the erupting flux rope and disappear as the flux rope rises. As these loops vanish from the AIA channels with cooler temperature response functions, plasma around the same location brightens in the hotter AIA 131 \AA{} channel, suggesting possible plasma heating.

\begin{figure*}[!htb]
 \centerline{\includegraphics[width=1.0\textwidth,clip=]{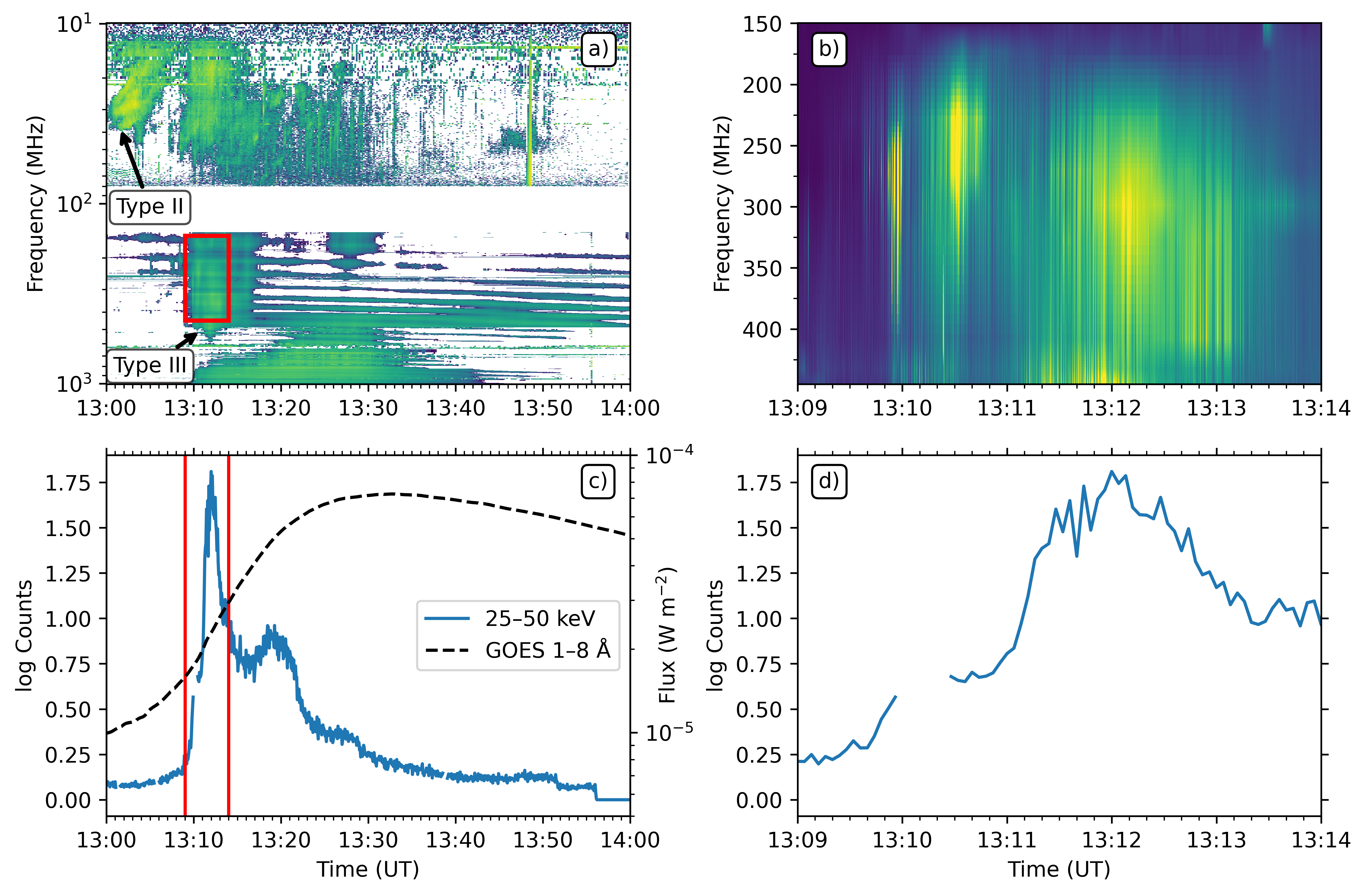}}
 \caption{a) Radio flux in the frequency range $10\textrm{--}1004\ \mathrm{MHz}$ from NDA and ORFEES. A type II burst is seen from 13:00 UT with a starting frequency of $30\ \mathrm{MHz}$. Strong type III bursts are observed from 13:09--13:14 UT. The red box corresponds to the time and frequency range represented in panel b.
 b) Zoom-in on the time and frequency range indicated by the red box in panel a, showing the radio flux around the time of peak hard X-ray activity.
 c) RHESSI hard X-ray light curve in the energy range $25\textrm{--}50/ \mathrm{keV}$ (solid blue line). Spikes in the data that were anomalously larger than neighbouring values have been removed. GOES disc-integrated soft X-ray emission is shown by the dashed black curve. The vertical red lines show the time interval presented in panels b and d.
 d) Zoom around the main peak of RHESSI hard X-ray emission from panel c.
 }
 \label{fig:secchirh}
 \end{figure*}

The flux rope CME drove a shock at its leading edge that caused a Type II radio burst from 12:55 UT \citep[][see Figure \ref{fig:secchirh}a]{palmerio2019ApJmultipoint}, and encountered the trailing edge of the previous, slower CME at 13:12 $\pm$ 00:09 UT \citep{ding2014interaction}. This shock-CME interaction has been attributed as the cause of moving radio bursts that were observed at frequencies of 150.9 MHz and 173.2 MHz by the NRH \citep{morosan2020interaction}. 

\subsection{The Flare}

In association with this flux rope CME, an M-class flare was detected in NOAA AR 11745 by GOES. There were no other significant flares on the solar disc around this time, so we can assume the disc-integrated soft X-ray flux measured by GOES is representative of the flare evolution seen in NOAA AR 11745. The soft X-ray counts increase gradually from the time the active region filament brightens and the arched flux rope begins to rise ($\approx$ 12:15 UT). 
A sharp increase in soft X-rays occurs at 13:08 UT (the registered start time of the GOES M5.0 flare), and the flare peaks at 13:32 UT before gradually decreasing in intensity over the next several hours. 

Two J-shaped flare ribbons brighten in the active region (Figure \ref{fig:aia_multi}b and \ref{fig:aia_multi}d) and separate from each other from 13:09--14:30 UT, signifying the expansion of the flaring arcade sweeping outward as magnetic reconnection continues at higher and higher altitudes beneath the erupting CME (\citealp{kopp_pneuman1976reconnection} and \textit{e.g.,} \citealp{fletcher2001ribbons}).
In the 131 \AA{} AIA channel, the flare arcade appears $\approx 100 ''$ ($\approx 70\ \mathrm{Mm}$) long.

\begin{figure*}
 \centerline{\includegraphics[width=0.8\textwidth,clip=]{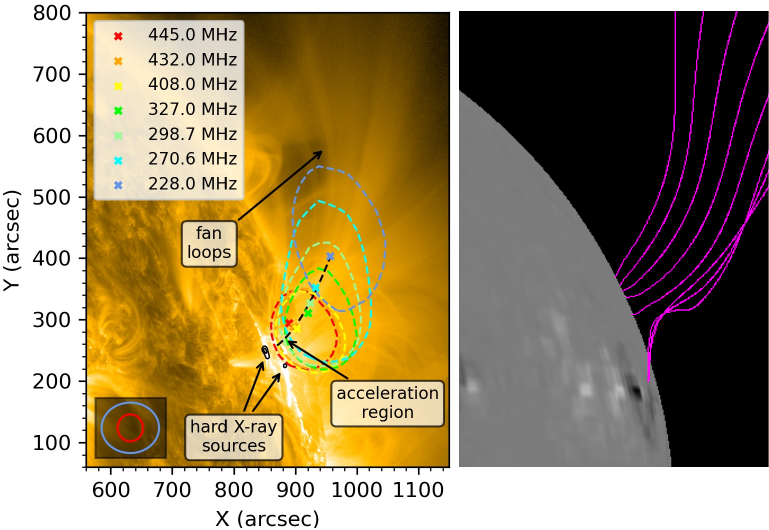}}
 \caption{Left: Fan loop structure seen in EUV AIA 171 \AA{} at 13:11:59 UT on 22 May 2013. Coloured contours of radio sources at different NRH frequencies are drawn at $70\%$ of the maximum intensity and their centroid coordinates (coloured crosses) are taken by fitting 2D Gaussian ellipses to each source. The (quadratic) beam path fitted to these centroids is shown by the black dashed line. The $70\%$ size of the NRH beam at $445\, \mathrm{MHz}$ and $228\, \mathrm{MHz}$ is shown in the bottom left. Hard X-ray sources observed by RHESSI are shown with black contours at $70\%$ and $30\%$ of the maximum intensity observed in the $27-70\, \mathrm{keV}$ band. Right: Potential field source surface extrapolation at 12:04 UT showing only open magnetic field lines (pink).}
 \label{fig:aia_nrh_pfss}
 \end{figure*}

In the AIA 171 \AA{} passband, we see fan loops that extend almost radially away from the north-west of NOAA AR 11745 (top panel of Figure \ref{fig:aia_nrh_pfss}). A PFSS extrapolation on 22 May shows open magnetic field rooted in this general location that aligns well with the observed fan loops (bottom panel of Figure \ref{fig:aia_nrh_pfss}).
We learn more about this fan structure by examining the evolution of the region over the previous several days (when it was closer to disc-centre). 
Using 171 \AA{} EUV images and another PFSS extrapolation, we can see that there were quasi-open EUV fan loops to the north west of NOAA AR 11745 when the region was close to disc centre on 16 May 2023, suggesting this aspect of the active region's structure is somewhat long-lived. 

\section{Electron beams}

\subsection{Confined electron beams} \label{sec:confined_beams}

Hard X-ray intensity in the $25\textrm{--}50\ \mathrm{keV}$ RHESSI band began to rise at 12:57 UT and peaked at 13:12 UT (see Figure \ref{fig:secchirh}c). There was also a second, lower peak at 13:19 UT. Two coherent hard X-ray sources in this energy range were imaged by RHESSI between 13:16:28 UT and 13:22:28 UT (Figure \ref{fig:aia_nrh_pfss}). The X-ray images were created using the Pixon algorithm with an integration time of 2 minutes.  These sources lie along the flare ribbons, indicating that they are signatures of energetic particles bombarding the solar surface after having been accelerated at a reconnection site in the corona and streaming down newly reconnected flaring loops.

\begin{figure}
 \centerline{\includegraphics[width=0.47\textwidth,clip=]{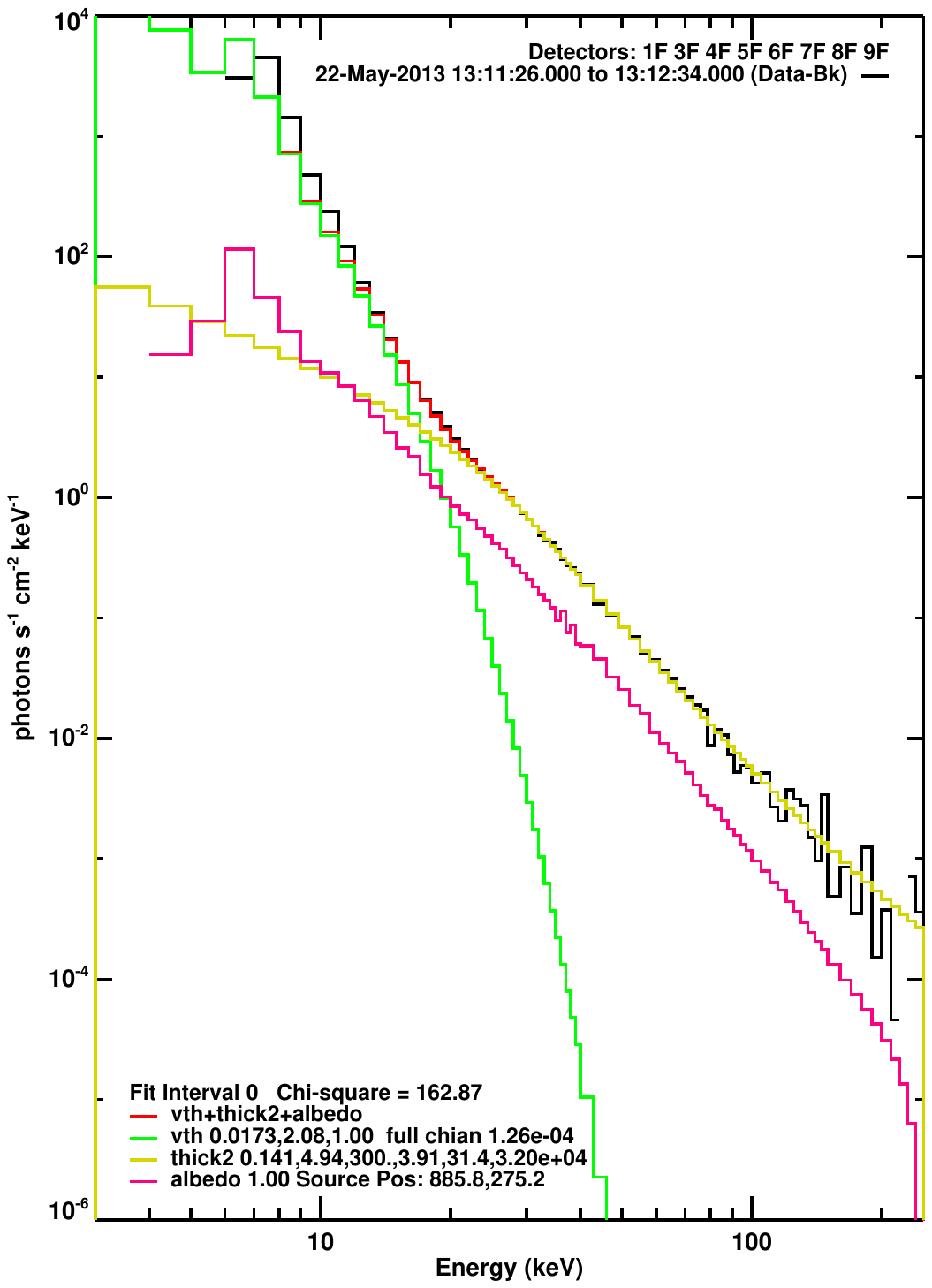}}
 \caption{X-ray spectrum fitted with a power-law, thermal distribution and correcting for albedo.}
 \label{fig:rhessi_fit}
 \end{figure}

We estimate the energy distribution of the downward accelerated particles from the hard X-ray sources.  We fit the main impulsive phase of the X-ray flare as observed by RHESSI (Figure \ref{fig:rhessi_fit}), which lasts from 13:11:26 to 13:12:34 UT (see Figure \ref{fig:secchirh}d).   The high energy part of the spectrum was fitted with the `thick 2' power-law component in the OSPEX software, taking into account corrections for albedo.  The low energy part of the spectrum was fit with a thermal component.  The spectral index fitting the power-law is found to be 5 in energy space ($10$ in velocity space) and the low energy cutoff is around 30 keV.  Fitting the spectra in 4-second intervals (RHESSI spin period) over the $\approx 1$ minute peak in hard X-rays, neither the spectral index or the low energy cutoff was found to vary during this interval.

\subsection{Escaping electron beams} \label{sec:radio}

We observe electron beams accelerated upward from the coronal reconnection site, away from the Sun, evidenced by type III radio bursts.  The radio bursts are first observed at relatively high frequencies by the NRH and ORFEES (corresponding to emission at low-coronal plasma densities), and then at lower frequencies, seen by the NDA and eventually as an interplanetary type III burst by Wind/WAVES.

The intense type III radio bursts are first seen in the NRH data at 13:09:54 UT. Many successive bursts occur roughly every two seconds until 13:13:09 UT, totalling more than $100$ bursts (for example, see Figure \ref{fig:secchirh}b).  Each burst is associated with a beam of electrons traversing a gradient of decreasing density as they escape the Sun.

Strong radio sources are imaged by the NRH at frequencies from $445.0\textrm{--}228.0\ \mathrm{MHz}$ (see Figure \ref{fig:aia_nrh_pfss}). 
There is also some emission at $173.2\ \mathrm{MHz}$ and $150.9\ \mathrm{MHz}$, but much weaker. 
Each source is offset slightly from the others along roughly the solar radial direction, with higher frequency sources closer to the Sun and lower frequency sources further from the surface. 
This resembles a column of radio emission and signifies the path of the escaping electron beams. 
The radio sources --- and therefore the electron beam paths --- closely follow the set of EUV fan loops described in Section \ref{sec:observations} that extend approximately radially outward from NOAA AR 11745 (see Figure \ref{fig:aia_nrh_pfss}). 

\subsubsection{Speed of escaping electron beams}

We estimate the speeds of several of the observed electron beams.
For each beam, we find the time of peak flux at each frequency by fitting Gaussian distributions to the measured flux-time profiles. 
Then, we select the NRH image that was taken closest to the peak flux time for each frequency and fit a 2D elliptical Gaussian to the observed radio source. We take the coordinate of the fitted maximum intensity of each ellipse as the centroid of the beam at that frequency and fit an approximate beam path to these coordinates (and a point at the base of the EUV fan loops).

The fitted source centroid of the 445 MHz appears slightly to the north east of the 432 MHz and 408 MHz sources (\textit{i.e.} further from the Sun).  
However, the sources are all relatively large, and much of the extended 445 MHz source still appears directly between the lower frequency sources and the flare arcade. Therefore, to avoid potentially overfitting, we use a quadratic function (rather than a cubic) to represent the curved path of the electron beam.

We take the points along the fitted beam path where the distance to each source centroid is minimal and use these as representative coordinates for each radio source along the fitted beam path. 
We calculate the plane-of-sky distances along the beam path between these points, and the differences between the peak flux times at each frequency. 
The gradient of the fit to these distances and times therefore gives an estimation of the plane-of-sky speed of each electron beam.  
We assume the type III was created by fundamental emission, on account of high polarisation measurements made by the NRH of $\sim 60\%$.

\begin{figure}[!t]
     \centering
     \begin{subfigure}
         \centering
         \includegraphics[width=0.47\textwidth]{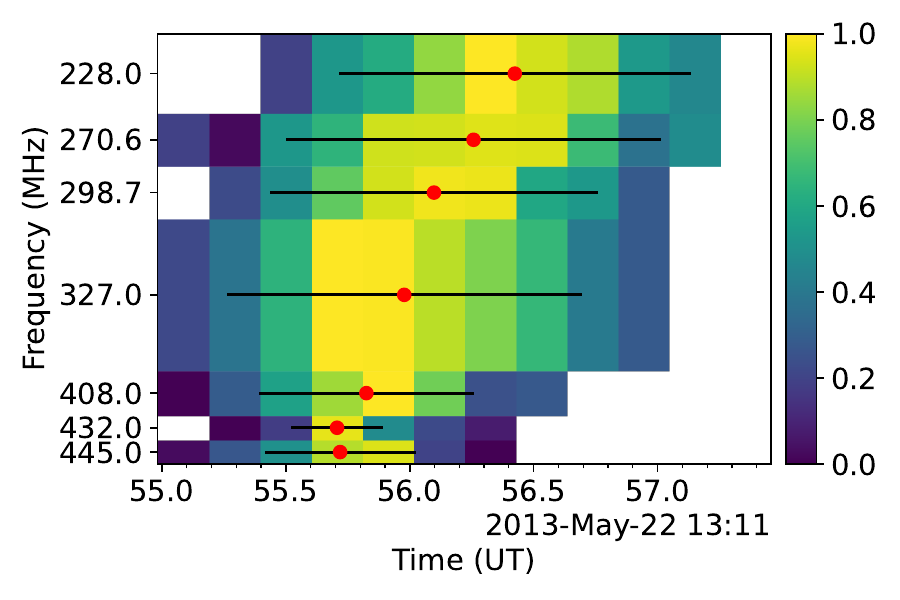}
     \end{subfigure}
     \begin{subfigure}
         \centering
         \includegraphics[width=0.45\textwidth]{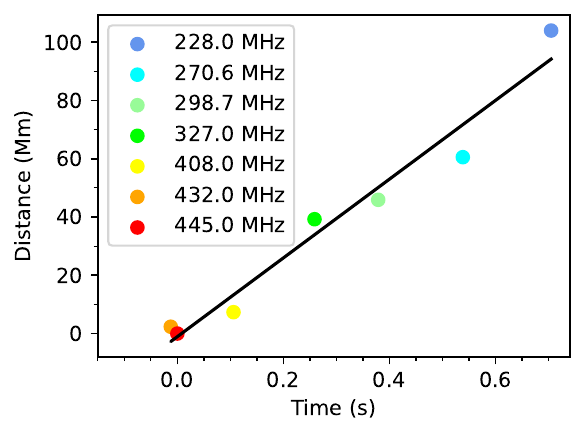}
     \end{subfigure}
        \caption{Top: Type III burst from around the RHESSI $25\textrm{--}50\, \mathrm{keV}$ peak time. Measurements associated with the previous and subsequent burst are removed and fluxes are normalised to the peak value at each each frequency. Times of peak flux at each frequency from Gaussian fitting are indicated with red circles, and full-width at half-maximum times are indicated by horizontal black lines. Bottom: Distance along the fitted beam path vs peak flux time for each frequency, relative to the $445\ \mathrm{MHz}$ source. The gradient of the fit represents the speed of the electron beam, which is $0.45c$.}
        \label{fig:nrh_beam_131154_speed_fit}
\end{figure}

We repeat the above analysis to determine the speeds of several electron beams that were observed between 13:10 UT and 13:14 UT, finding speeds that range from $0.44\textrm{--}0.59c$ (frequency drift rates $\approx{}300\rm{-}400~\rm{MHz}~\rm{s}^{-1}$). 
However, the fastest of these speeds are found at times when there are multiple beams in particularly quick succession, which led to increased levels of background flux and less reliable fits.
Therefore, we are less confident in our estimation of these higher speeds.
In the rest of this section, we analyse a beam that we confidently estimate to have had a plane-of-sky speed of $0.45c$ (see Figure \ref{fig:nrh_beam_131154_speed_fit}). 
The associated type III burst was first observed in the $445\ \mathrm{MHz}$ NRH channel at 13:11:56 UT, and was therefore accelerated around the time of the peak $25\textrm{--}50 \mathrm{keV}$ hard X-ray emission measured by RHESSI. 

\subsubsection{Altitude of the acceleration region}

\begin{figure}
 \centerline{\includegraphics[width=0.45\textwidth,clip=]{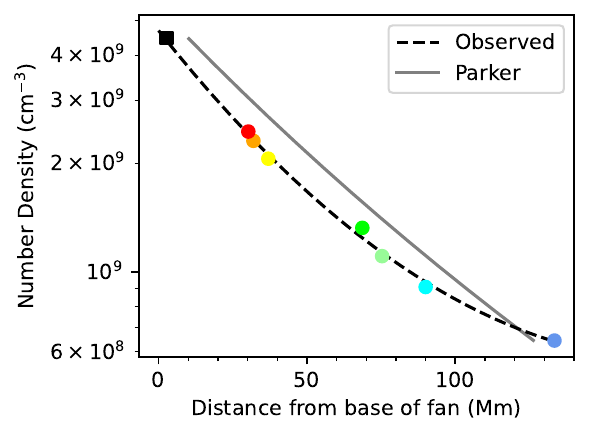}}
 \caption{Density profile with distance from the base of the fan loops fitted to the observed radio sources (coloured circles matching the frequencies in other Figures.). The black square represents a 600 MHZ source along the fitted beam path ($\approx 3\, \mathrm{Mm}$ from the base of the fan loops). The density at the base of the fan loops is $\num{4.8e9}\, \mathrm{cm}^{-3}$ ($\approx 620\ \mathrm{MHz}$). For comparison, the Parker density profile used in our simulation is indicated by the solid grey line.}

 \label{fig:density_profile}
 \end{figure}

We use the observed radio sources to estimate how the local plasma density varies with height in the corona, and thereby infer the density (and associated plasma frequency) around the acceleration region.
Assuming the radio sources are produced by fundamental plasma emission, we calculate the electron density $n_e$ of the emitting plasma in each source, as
\begin{equation}
n_e = \frac{4\pi^2m_e\epsilon_0}{q_e^2} f^2,
\label{eqn:density_freq}
\end{equation}
where $f$ is the radio source frequency.
By estimating a quadratic fit to the logarithm of plasma density and the distance of the radio source centroids along the beam path from the base of the fan loops, we derive an electron density profile along the beam (see Figure \ref{fig:density_profile}). 

The inferred coronal density profile suggests a plasma density at the base of the fan loops of $\num{4.8e9}\, \mathrm{cm}^{-3}$, which corresponds to a plasma frequency of $\approx 620\, \mathrm{MHz}$.  We use ORFEES data to identify the time and frequency at which the radio emission began (\textit{i.e.} after the beam had travelled the instability distance from the acceleration region).  The type III starting frequency exhibits the characteristic low-high-low behaviour reported by \citep{reid2014starting}. At the beginning and the end of the period of type III bursts (\textit{i.e.}, at 13:09 UT and 13:14 UT), the starting frequency is $\approx 485\ \mathrm{MHz}$, but around the time of peak hard X-ray activity (13:12 UT), the starting frequency is $\approx 600\ \mathrm{MHz}$ (see Figure \ref{fig:secchirh}a).  Assuming density, and therefore frequency, indeed simply decrease with distance from the Sun, the plasma frequency in the acceleration region would be lower than at the the base of the fan loops, but greater than the type III starting frequency of $\approx 600\ \mathrm{MHz}$ observed by ORFEES.  For illustration, a $600\, \mathrm{MHz}$ source corresponds to a plasma density of $\num{4.5e9}\, \mathrm{cm}^{-3}$ and would lie at a distance of $\approx 3\, \mathrm{Mm}$ from the base of the fan loops along the fitted beam path.

There is only a $\approx 10\, \textrm{Mm}$ height range (in the plane-of-sky) between the height where the type III bursts begin and the flaring loops where the acceleration region must lie. Given the spectral index, $\alpha=9.4$ from hard X-ray observations, we can use the Equation \ref{eqn:instability_distance} to deduce the acceleration region length must be $d \sim 1\, \textrm{Mm}$.

\subsubsection{Cross-section of the acceleration region}

\begin{figure}
 \centerline{\includegraphics[width=0.5\textwidth,clip=]{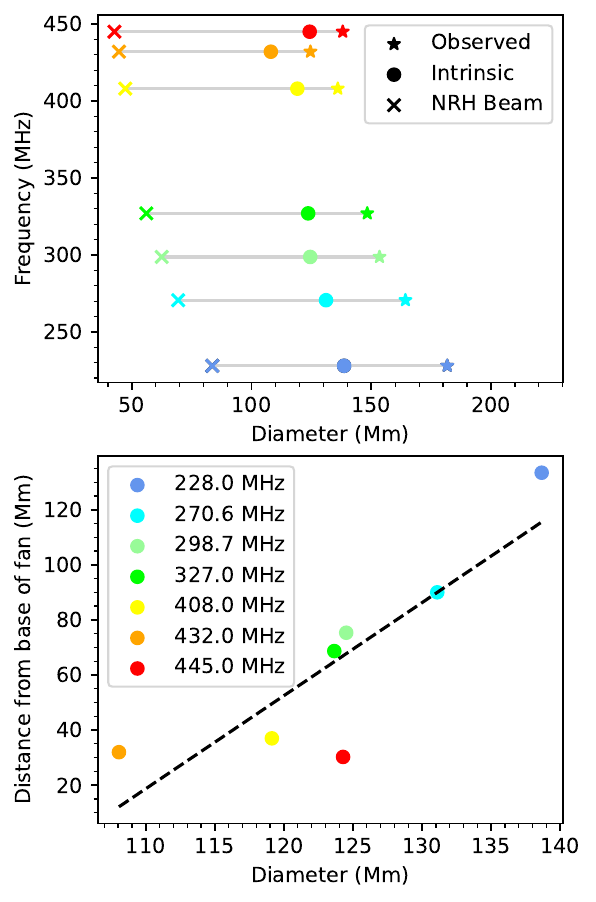}}
 \caption{Top: FWHM diameters of ellipses fitted to observed radio sources (stars), intrinsic source FWHM after accounting for frequency-dependant scattering (circles), and the NRH beam diameter (crosses) at each frequency. Bottom: Intrinsic source FWHMs from the top panel plotted against the distance of each source centroid along the fitted beam path. The fit shown by the dashed line is $y = 3.4x - 353$.}
 \label{fig:diameter_vs_distance}
 \end{figure}

Figure \ref{fig:diameter_vs_distance} shows how the beam diameter varies with increasing distance along the beam path.   Here, we define the beam diameter, $D = 2 \sqrt{2\ln{2}\ \sigma_a \sigma_b}$, where $\sigma_a$ and $\sigma_b$ are the major and minor standard deviations of the ellipses fitted to the radio sources at each frequency. Simply, this diameter is the average FWHM intensity of each elliptical source. 

We account for the effects of frequency-dependant scattering to estimate the intrinsic source sizes from the observed sizes \citep{kontar2019scattering}.
The 445 MHz source has an observed FWHM of 138 Mm, and based on the relationship from \citet{kontar2019scattering} that a 35 MHz source has a FWHM from scattering of $1.1\, R_{\odot}$, the scattering FWHM at 445 MHz is $60\, \mathrm{Mm}$. Subtracting this in quadrature from the observed FWHM gives an intrinsic source size of 124 Mm at 445 MHz.
For reference, the NRH beam size at 445 MHz is 43 Mm.

Generally, we see a linear expansion of the source diameters along the beam (see Figure \ref{fig:diameter_vs_distance}). 
Unusually, the 445 MHz source seems larger than the 432 MHz and 408 MHz sources, but this may simply result from poor fitting to the 445 MHz source.
The smallest radio source (432 MHz) has an intrinsic diameter of $\approx 111\ \mathrm{Mm}$, and the linear fit to the beam diameters in frequency ($d = 3.4D - 353\, \mathrm{Mm}$, where $d$ is the distance from the base of the fan loops) suggests the beam would have had a diameter of $104\, \mathrm{Mm}$ at the base of the fan loops, if it had originated there and expanded at a constant, linear rate. 
We expect the beam originated from an acceleration region somewhere between the base of the fan loops and the height at which it started to produce radio emission, suggesting the beam had a diameter $100<r<110\ \mathrm{Mm}$ at its origin. 
However, it is plausible that the beam could have started smaller than this, first undergoing a rapid, nonlinear expansion before it began producing the radio sources observed by the NRH. 

\subsection{Electron beam energy density} \label{sec:simulation}

To estimate the energy distribution of the accelerated particles in the escaping beams, we can make the assumption that the distributions of the upward and downward accelerated electron beams are the same.  
Previous studies have shown a correlation between these energy distributions \citep{krucker2007electron,dresing2021electron} using X-rays and in situ electron measurements. 
In Section \ref{sec:confined_beams}, we fitted the main impulsive phase of the X-ray flare and found a spectral index of $10$ in velocity space, with a low energy cutoff around 30 keV.

\citet{reid_kontar2018} showed a direct dependence of the type III velocity on the spectral index, the background density, and the density of the accelerated electrons. They simulated an electron beam with the initial source function:

\begin{equation} \label{eqn:t_b_orig}
S(v,r,t) = g(v)\exp{\left(\frac{(-x-x_0)^2}{d^2}\right)} A_t\exp{\left(\frac{(-t-t_0)^2}{\tau^2}\right)},
\end{equation}
where $\tau$ and $d$ describe the characteristic temporal and spatial properties of the acceleration region.  The velocity dependence $g(v)$ was a single power-law (spl) that we denote here as $g_{\rm{spl}}(v)$, given by the following:

\begin{equation}
    g_{\rm{spl}}(v) = A_{\rm{spl}}\, v^{-\alpha},
\end{equation}

with

\begin{equation}
    A_{\rm{spl}} = \frac{n_{\rm beam}(\alpha-1)}{v_{\rm min}^{1-\alpha} - v_{\rm max}^{1-\alpha}},
\end{equation}

for the beam density $n_{\rm beam}$, minimum and maximum velocities $v_{\rm min}$ and $v_{\rm max}$ corresponding to energies of 0.3 keV and 125 keV, respectively. Above $v_{\rm{max}}$, minimal Langmuir waves are generated by the electron beam on account of low electron energy densities and their consideration would require accounting for relativistic effects.

Using the background density model estimated from the radio emission (Figure \ref{fig:density_profile}) and the spectral index obtained from the X-ray observations, we can find the density of the accelerated electrons that accurately simulates the propagation of the observed electron beam velocity. Figure 9 in \citet{reid_kontar2018} shows that, for similar peak velocities, increasing the spectral index by 1 corresponds to an order of magnitude increase in beam density. For example, at a peak velocity of $\approx{}0.5c$, beams with spectral indices of $\alpha=6$ and $\alpha=7$ have beam densities of $n_{\rm beam}=10^{7}\, \rm{cm}^{-3}$ and $n_{\rm beam}=10^{8}\, \rm{cm}^{-3}$, respectively. Since our fundamental type III emission drifts with a comparable velocity of 0.45c, the observed spectral index of $\alpha=10$ suggests a beam density of $n_{\rm beam}=10^{11}\, \rm{cm}^{-3}$. 

A beam density of $n_{\rm beam}=10^{11}\, \rm{cm}^{-3}$ would result in a nonphysical density given the coronal acceleration region obtained from radio observations has a background electron density around $4\times 10^9~\rm{cm}^{-3}$ (plasma frequency around 600 MHz).  Consequently, we do not expect the power-law of accelerated particles to extend all the way down to 0.3 keV.  Instead, we assume a broken power-law (bpl) with a break around 30 keV (the break energy is derived from the X-ray observations).  The rationale is that Coulomb collisions would damp the acceleration of electrons below 30 keV.  

For the broken power-law, we define the velocity dependence 
\begin{eqnarray}\label{eqn:bpl}
g_{\rm{bpl}}(v < v_0)  &= A_{\rm{bpl}} \quad \quad 
g_{\rm{bpl}}(v\geq v_0)  &= A_{\rm{bpl}} \left(\frac{v}{v_0}\right)^{-\alpha} 
\end{eqnarray}
where
\begin{equation}
\begin{aligned}
A_{\rm{bpl}} = n_{\rm beam} \left[ v_0-v_{\rm min}+\frac{v_{\rm max}(v_{\rm max}/v_0)^{-\alpha} - v_0}{1-\alpha}\right]^{-1}
\end{aligned}
\end{equation}
is normalised so that $n_{\rm beam}$ describes the beam density above 30 keV (velocity $v_0$ around $10^{10}~\rm{cm~s}^{-1}$).  The broken power-law was investigated in \citet{reid_kontar2018} but for smaller break energies.

\begin{center}
\begin{table*}
\centering
\caption{Initial parameters for the electron beam injected into the solar corona.}
\begin{tabular}{ c  c  c  c  c  c  c }
\hline\hline
\multirow{2}{*}{Energy Range} & \multirow{2}{*}{Break Energy} & Spectral Index & \multirow{2}{*}{Injection Time} & \multirow{2}{*}{Beam Size} & \multirow{2}{*}{Source Height} & Beam density \\
 & & above break & & & & above break
\\ \hline
$0.28-125$~keV &  $30$~keV & $\alpha=10.0$ & $\tau=0.001$~s & d=$10^8$~cm & h=$15$~Mm &  $10^{2.5}~\rm{cm}^{-3}$  \\
\hline
\end{tabular}
\vspace{20pt}
\label{tab:beam_sun}
\end{table*}
\end{center}

To test this prediction of beam density, we simulated an electron beam being injected into the corona with source function described by Equation \ref{eqn:t_b_orig} that has an electron velocity broken power-law described by Equation \ref{eqn:bpl}.  The kinetic equations used to propagate the system through time are described in \citep{reidkontar2017b,reid_kontar2018}.  The Parker density model \citep{parker1958density} is used, scaled to provide an acceleration region around 15 Mm above the solar surface, with a background frequency of 600 MHz.  This density model naturally provides a similar background frequency profile that matches the distances derived from the radio observations (Figure \ref{fig:density_profile}).

\begin{figure}
 \centerline{\includegraphics[trim=50 50 50 0,clip,width=0.45\textwidth]{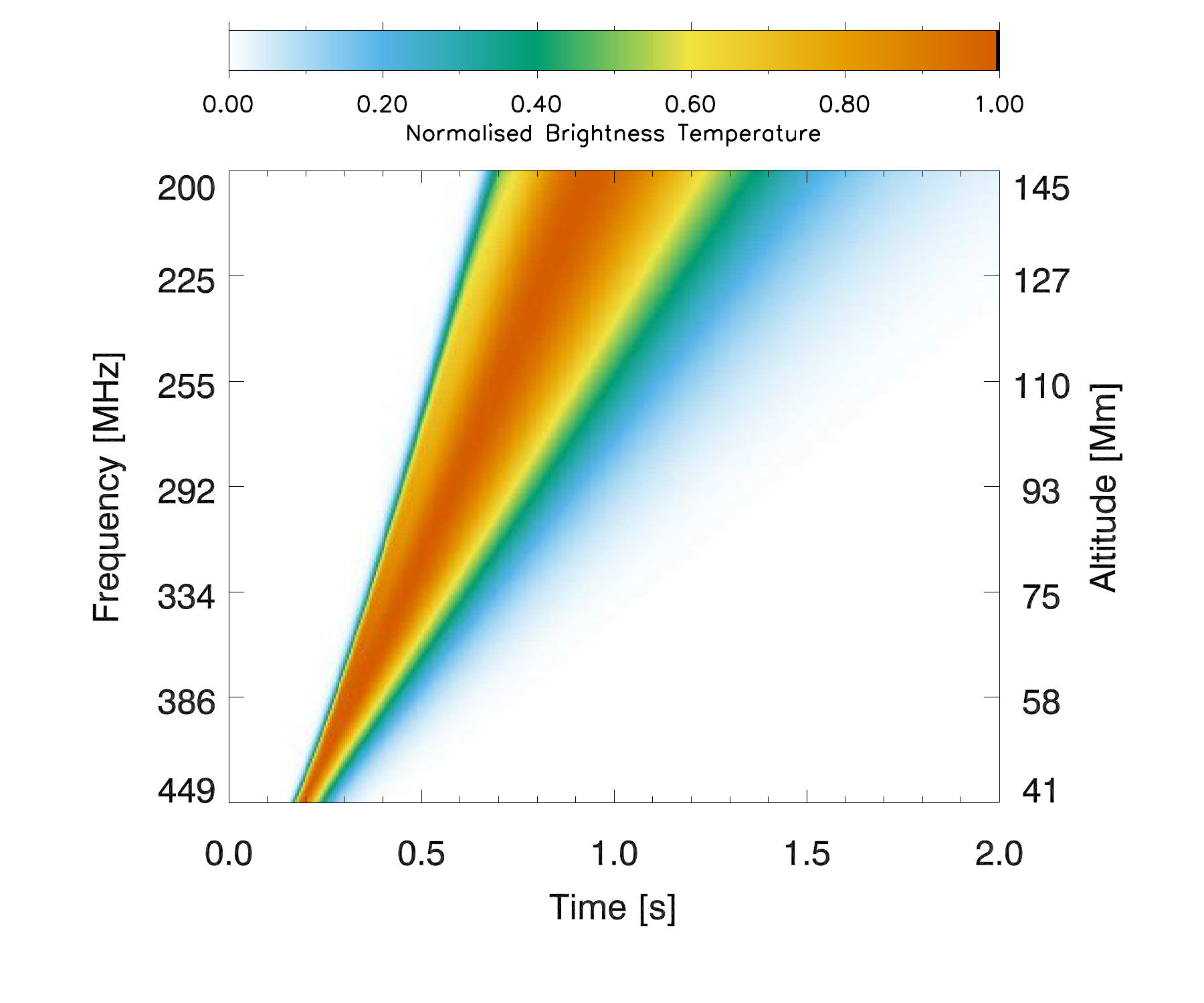}} 
 \caption{Simulated dynamic spectrum using the electron beam parameters derived from the radio and X-ray observations.}
 \label{fig:simulation}
 \end{figure}

To reproduce the electron beam described above that we observed around the peak of $25-50\ \mathrm{keV}$ hard X-ray activity, as measured by RHESSI, we simulated an electron beam with a spectral index of 10 (in velocity space) with an initial electron density of $n_{\rm beam} \sim 10^{2.5}\ \mathrm{cm}^{-3}$ for $E > 30\ \mathrm{keV}$. For $E < 30\ \mathrm{keV}$, we use a flat spectrum.  The initial injection properties are given in Table \ref{tab:beam_sun}.

We infer a magnetic field expansion angle of $17^{\circ}$ from the expansion of the intrinsic source sizes shown in the bottom panel of Figure \ref{fig:diameter_vs_distance}. This is lower than expansion angles used in other works to explain observed fluxes at 1 AU ($33.6^{\circ}$; \citealp{reid2010electrons,reid2013electrons}).

The resultant simulated electron beam (shown in Figure \ref{fig:simulation}) had a velocity of $0.45c$, matching the velocity of the beam we obtained by fitting to NRH observations (Figure \ref{fig:nrh_beam_131154_speed_fit}).

We can analytically calculate the energy density of the beam by assuming a single power-law. The energy density of the beam above the cutoff energy is:

\begin{equation}
\begin{aligned}
    \epsilon{} & = \frac{1}{2}m_{e} \int_{v_{0}}^{v_{max}} g(v \geq v_{0}) \, v^{2}\, dv \\
               & = \frac{1}{2} \frac{m_{e}A_{v}}{3-\alpha} \left[v^{3-\alpha}\right]_{v_{0}}^{v_{max}}.
\end{aligned}
\end{equation}

For $v_0 = 10^{10}\, \mathrm{cm}\, \mathrm{s}^{-1}$, $v_{max} = 2\times10^{10}\, \mathrm{cm}\, \mathrm{s}^{-1}$ ($125\, \mathrm{keV}$), $n_b = 10^{2.5}\, \mathrm{cm}^{-3}$, and $\alpha=10$, we find $\epsilon = 2\times10^{-5}\, \mathrm{erg\, cm^{-3}}$.

\section{Discussion} \label{sec:discussion}

A CME erupted from NOAA AR 11745 at around 13:00 UT on 22 May 2023. An associated M5.0 flare began in the active region at 13:08 UT and peaked at 13:32 UT. RHESSI measured a steep rise in hard X-ray activity in the $25\textrm{--}50\ \mathrm{keV}$ range starting at $\sim$13:09 UT, with a strong peak at 13:12 UT and a second, smaller peak at 13:19 UT (see Figure \ref{fig:secchirh}c). Strong type III radio bursts started just before 13:10 UT and continued until shortly after 13:13 UT, with a burst occurring approximately every two seconds (Figure \ref{fig:secchirh}a and \ref{fig:secchirh}b). After this time, type III radio bursts continued to occur, but with much weaker fluxes. We suggest the rapid sequence of electron beams that caused the type III radio bursts were accelerated by quasi-periodic oscillatory magnetic reconnection \citep[\textit{e.g.}][]{mclaughlin2009oscillatory} associated with the M5.0 flare and the hard X-ray emission. Such quasi-periodic reconnection can be driven by an aperiodic driver \citep{mclaughlin2012bperiodicity}, such as magnetic flux emergence \citep{murray2009oscillatory,mclaughlin2012ageneration}, or perhaps in our event, the continuous inflow of magnetic field that would have occurred beneath the erupting hot magnetic flux rope.

\begin{figure*}
 \centerline{\includegraphics[width=1.0\textwidth,clip=]{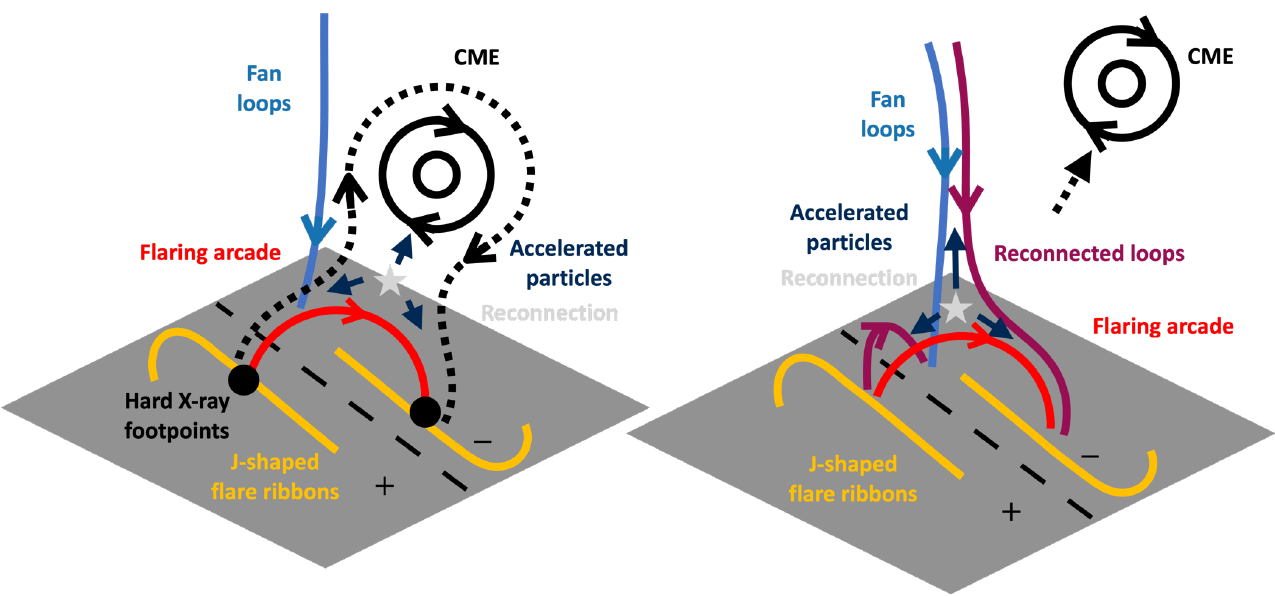}}
 \caption{Two-stage reconnection scenario. First, magnetic reconnection associated with a CME accelerates particles (left), and then further reconnection between either the flaring arcade or the overlying loops above the CME and quasi-open fan loops enables the particles to escape into interplanetary space.}
 \label{fig:cartoon}
 \end{figure*}

Radio imaging spectroscopy by the NRH shows that the radio bursts travelled along a set of fan loops rooted in the north western part of the active region. These loops are seen clearly in the EUV 171 \AA{} channel of AIA, and a PFSS extrapolation supports that the fan loops are open to the heliosphere (Figure \ref{fig:aia_nrh_pfss}). For the electrons accelerated in the core of the active region to escape along these quasi-open fan loops, a reconfiguration of the magnetic field must have occurred. 
We propose that either overlying loops that were draped over the erupting flux rope or the growing flare arcade itself expanded and pushed against the fan loops at the active region boundary, leading to component magnetic reconnection that enabled the beams of particles accelerated in the eruptive flare to access the quasi-open magnetic field and escape the Sun (see Figure \ref{fig:cartoon}). 
This is similar to the scenario simulated by \citet{masson2013escape,masson2019ApJescape}.

\subsection{Electron beam energy budget}

We have estimated the energy density in the acceleration region by inferring the electron beam speed and deducing the electron density via simulation.
We observe $\gtrsim \num{e2}$ strong type III bursts, so we can estimate the total energy budget available to accelerate the electron beams if we can constrain the volume of the acceleration region.
As shown in Equation \ref{eqn:instability_distance}, the length of the acceleration region is related to the instability distance between the acceleration region and the location where type III radio emission starts \citep{reid2011acceleration}.
Therefore, we can begin to constrain the size of the acceleration region by examining the distance the electrons travelled before they started emitting radio bursts.

The acceleration region should be above the observed flaring arcade and below the highest frequency ($445\ \mathrm{MHz}$) NRH source. 
The plane-of-sky distance between the $445\ \mathrm{MHz}$ fitted source centroid and the top of the flaring loops observed in AIA 171 \AA{} images is $\approx 15\ \mathrm{Mm}$. 
Considering only that the fan loops along which the radio sources follow are in the background of the active region, this plane-of-sky distance is an underestimated projected distance. 
On the other hand, a number of factors suggest the distance between the flare arcade and the acceleration region may be even smaller than it first appears, including that the scattering of the radio sources would shift the observed sources closer to the Sun \citep{kontar2019scattering}, the flaring arcade appears larger in other AIA channels than as we have considered in the 171 \AA{} channel (\textit{e.g.}, in particular the 131 \AA{} channel), and the ORFEES observations suggest the type III bursts start at $\approx 600\ \mathrm{MHz}$ around the time of peak hard X-ray activity, which would be relatively low in the corona. 
Whilst it is difficult to quantify the height of the acceleration region precisely, it is clear that the instability distance ($d\alpha$) between the acceleration region ($h_{\mathrm{acc}}$) and the height where the electron beams start producing radio emission ($h_{\mathrm{III}}$) must be very small (see Equation \ref{eqn:instability_distance}). 
Since the spectral index, $\alpha \approx 10$, even an instability distance $h_{\mathrm{III}} - h_{\mathrm{acc}} \approx 10\ \mathrm{Mm}$ suggests a very short acceleration region $\sim 1\ \mathrm{Mm}$.

To estimate the width of the acceleration region, we consider the magnetic reconnection scenario. It is thought that 3D magnetic reconnection occurs in quasi-separatrix layers (QSLs; \citealp{priest1995qsl,demoulin1996qsl}), so perhaps energetic particle acceleration also occurs in these layers. 
In our scenario, such a QSL would overlie the flaring arcade, and therefore the width of the acceleration region may be comparable to the width of the flaring loops and the fan loops (just above the flaring arcade) observed in the EUV images.

Taking the width of the acceleration region as comparable to the width of the EUV fan loops along which the electron beams escape ($\approx 15\ \mathrm{Mm}$ just above the flare arcade, similar to the width identified by \citealp{guo2012accelregion}) up to the full width of the flaring arcade itself ($\approx 70\ \mathrm{Mm}$; as in the scenario of a loop top acceleration region) or the $\approx 100\, \mathrm{Mm}$ inferred by the beam expansion (see Figure \ref{fig:diameter_vs_distance}, the acceleration volume would be $\num{e26}\textrm{--}\num{e28}\ \mathrm{cm}^{3}$ (although uncertainty is introduced here by assuming the cross sectional area of the region is equal to the width$^{2}$).
This compares favourably with the cylindrical volume found by \citeauthor{gordovskyy2020acceleration} (\citeyear{gordovskyy2020acceleration}; $10^{26}\, \mathrm{cm}^{3}$) and the loop top energy release region volume inferred by \citeauthor{fleishman2022volume} (\citeyear{fleishman2022volume}; $10^{27}\, \mathrm{cm}^{3}$).

Considering  $\sim{}10^{2}$ beams were accelerated, a volume in this range with an electron density above $30\ \mathrm{keV}$ of $n_{beam} \sim 10^{2.5}\, \mathrm{cm}^{-3}$ and energy density $\epsilon_{beam} \approx \num{2e-5}\, \mathrm{erg\ cm}^{-3}$ (supported by our simulation), would contain a total of $\num{e31}\textrm{--}\num{e33}$ electrons and an energy of $10^{23}\textrm{--}10^{25}\ \mathrm{erg}$.
These numbers of escaping electrons are comparable to those found by \citeauthor{krucker2007electron} (\citeyear{krucker2007electron}; $10^{31}\textrm{--}10^{33}$), and are at the higher ends of the numbers found by \citeauthor{james2017electron} (\citeyear{james2017electron}; $10^{30}\textrm{--}10^{32}$) and \citeauthor{dresing2021electron} (\citeyear{dresing2021electron}; $10^{30}\textrm{--}10^{31}$).
Our energy estimates are comparable to those of \citeauthor{james2017electron} (\citeyear{james2017electron}; $10^{23}\textrm{--}10^{25}\ \mathrm{erg}$).

\section{Conclusions} \label{sec:conclusions}

On 22 May 2013, an M5.0 flare occurred in NOAA AR 11745 after a flux rope CME erupted.
RHESSI measured a strong peak of $25\textrm{--}50\ \mathrm{keV}$ hard X-ray emissions and imaged sources at the footpoints of flaring loops in the AR core.
More than \num{e2} strong type III radio bursts were observed in a period of less than 5 minutes, suggesting many electron beams were being accelerated from the flare site. 
Radio sources in the range $445.0\textrm{--}228.0\ \mathrm{MHz}$ aligned with quasi-open fan loops at the northwestern edge of the active region.
We propose that component magnetic reconnection between either the expanding loops above the erupting flux rope or the growing flare arcade beneath erupting material and quasi-open fan loops at the periphery of the active region enabled the flare-accelerated electrons to escape into interplanetary space.

We used radio imaging observations to determine the speeds of several electron beams, finding high speeds that range from $0.44\textrm{--}0.59c$. The beam closest in time to the hard X-ray peak had a speed of $0.45c$, and the spectral index at this time was $\alpha = 10$.
We successfully simulated the $0.45c$ electron beam using this spectral index and an electron density above $30\ \mathrm{keV}$ in the source region of $10^{2.5}\, \mathrm{cm}^{-3}$, giving insight into the conditions in the acceleration region.

The acceleration region should be located above the top of the flare arcade and below the height at which the radio bursts begin. 
The plane-of-sky distance between the $445\ \mathrm{MHz}$ source centroid and the top of the flaring loops observed in 171 \AA{} AIA images is $\approx 15\ \mathrm{Mm}$, but the type III bursts are observed by ORFEES to start at $\approx 600\ \mathrm{MHz}$, which would occur very close to the flare arcade. 
Furthermore, the electron beams must travel an instability distance before they start producing radio emission, which is related to the spectral index and the length of the acceleration region. 
Our findings suggest this instability distance must be very short, and since the measured hard X-ray spectral index is large, the acceleration region must be $\sim 1\ \mathrm{Mm}$.

A radially short but tangentially wide acceleration region above the flaring arcade with a width comparable to the size of the fan loops or the flaring arcade ($15\textrm{--}100\, \mathrm{Mm}$) could contain $\num{e31}\textrm{--}\num{e33}$ electrons and a total energy of $10^{23}\textrm{--}\num{e25}\ \mathrm{erg}$. 
We suggest quasi periodic reconnection accelerated the many observed electron beams from a thin but broad QSL atop the flare arcade.

Future work could compare estimates of escaping particle beam properties deduced from remote sensing observations --- as we have made for the first time in this work --- with in situ measurements taken by \textit{Solar Orbiter} \citep{mueller2020Orbiter} within the inner helisophere. This will require magnetic connectivity to be established between particle acceleration regions on the Sun and the spacecraft. \\

A.W.J. and H.A.S.R. acknowledge funding from the STFC Consolidated Grant ST/W001004/1.
We thank the radio monitoring service at LESIA (Observatoire de Paris) to provide value-added data that have been used for this study.
We thank the RSDB service at LESIA / USN (Observatoire de Paris) for making the NRH/ORFEES/NDA data available. 
We acknowledge the use of the RHESSI Mission Archive available at \url{https://hesperia.gsfc.nasa.gov/rhessi/mission-archive}.
Additional data courtesy of NASA/SDO and the AIA and HMI science teams.
SOHO is a project of international cooperation between ESA and NASA.
STEREO is the third mission in NASA’s Solar Terrestrial Probes program.

%

\vspace{5mm}
\facilities{NRH, ORFEES, NDA, RHESSI, GOES, SDO(AIA and HMI), SOHO(LASCO), STEREO(SECCHI)}


\software{SunPy (version 4.0.5; \citealp{sunpy_community2020,sunpy_4.0.5}),
          Astropy (version 5.0.4; \citealp{astropy2013,astropy2018v2.0,astropy2022v5.0}).
          }







\bibliography{bibliography}{}
\bibliographystyle{aasjournal}



\end{document}